\documentstyle[12pt]{article}

\begin{document}

\title{Arbitrariness in defining fractal basins:
Relations between open and closed systems}

\author{ {\it A. E. Motter$^a$ and P. S. Letelier$^b$}\\
Departamento de Matem\'atica Aplicada - IMECC \\
Universidade Estadual de Campinas, Unicamp\\
13083-970 Campinas, SP, Brazil}

\date{}
\maketitle

\begin{abstract}

A discussion about
dependences of the (fractal) basin boundary dimension
with the definition of the basins and the {\it size}
of the exits is presented for systems
with one or more exits. In particular,
it is shown that the dimension is largely independent of
the choice of the basins, and decreases with the size
of the exits. Considering the limit of small exits,
a strong relation between fractals in exit systems and
chaos in closed systems is found. The discussion is
illustrated by simple examples of one-dimensional maps.

\end{abstract}

\vskip 1.0truecm

PACS numbers: 05.45.-a, 05.45.Df

Keywords: chaos, chaotic scattering, fractal, invariant set. 

$^a$E-mail: motter@ime.unicamp.br.

$^b$E-mail: letelier@ime.unicamp.br.

\newpage

\section{Introduction}

Fractals have been largely applied to characterize the
dynamics of systems with multiple modes of exit \cite{ott}.
For instance,
in the capture of orbits by attractors \cite{greb},
in the scattering of particles by located \cite{chen}
and by nonlocated potentials \cite{moura},
and in the initial-condition sensitivities caused
by chaotic transients \cite{bleher}.
In this context, fractal sets
are: Boundaries among basins of initial conditions
going out via different exits, singularities of scattering
functions, or singularities of exit-time functions.

An exit system has in general either one of the
two following kinds of behavior:
a) The system is stable,
which is the open counterpart of nonchaotic
closed systems, or b) the system presents some
sensitive dependence on initial conditions, the
open analogous of chaotic closed systems. The behavior
exhibited by the system is
determined by the nature of the nonattracting
invariant set of trapped orbits.
More precisely, it is through the dimension of the future invariant
set that the sensitivity of the system is quantified \cite{ott}.
This dimension can go from zero (no instabilities) to
the energy surface dimension (maximal sensitivity), relating the
cause of the sensitivity (invariant sets) with the phenomenon itself
(fractal structures in basin boundaries, exit-times, etc).

The aim of this paper is to discuss relations between different
fractal sets, the dependence of the dimension of the basin
boundaries with the definition of the basins (equivalently,
the definition of the exits), and the dynamical consequences
of fractals in closed (nonexit) systems.
Among the motivations of this work are the so far
unanswered questions:
(1) How arbitrary can be the definition of the basins?
A basin boundary gives information about the system when
its dimension reveals the dimension of the future invariant
set. Since the future invariant set and the basin boundary
are in general non identical sets, conditions on the definition
of the exits that ensure the equality of the dimensions have
to be found. 
(2) How to study a system whose orbits ``choose''
the exits only after the time evolution of the
initial conditions by arbitrarily large periods of time?
As an example consider the orbits of
light traveling around a periodically perturbed
black hole \cite{chan}.
By physical considerations, almost
all orbits starting on a Poincar\'e section
are (eventually) captured by the black hole
or scattered to infinity. Suppose that we are
interested in measuring the sensitivity of the
system for these two exits by computing the
dimension of the invariant orbits on the
Poincar\'e section.
For some initial conditions, the trajectory
integrated on a finite period of time goes unequivocally
to the black hole. Other orbits, however, remain
around the black hole for a long time
and it is in general difficult to determine
whether the final state will be at the black
hole or at infinity. This problem is a consequence
of our inability in integrating the system for an
infinite interval of time.
(3) What sort of relation communicates fractals in exit
systems with chaos in nonexit systems?
Fractal structures and chaos present a number of analogous
implications. For instance, both chaos and fractals prevent
integrability. Accordingly, it worth looking for
a common mechanism determining fractals in open systems
and chaos in closed systems.

In order to answer these questions we
restrict the discussion to a class of autonomous systems
for which almost all orbits of interest eventually exit,
and the exits are defined regions of the phase space.
For these systems, we shall consider each $N$-dimensional
manifestly invariant part of the phase space
(e.g., each energy surface), where $N$ is refered as
the effective phase space dimension.

In Section 2 we observe that the basin boundary dimension
is largely independent of the choices of the basins
if the {\it total} exit is fixed.
In this context we also discuss relations between the
dimensions of different fractal sets for 
different concepts of dimension. 
The dimension decreases, however, when the total
exit is enlarged.
The enlargement of the exits as a technique to study
systems whose capture of orbits by the
exits takes arbitrarily long time is considered in
Section 3.
The limit of small exits, on the other hand,
reveals the chaotic or regular character of
the corresponding closed system. 
Relations between fractals in exit systems and
chaos in the associated nonexit system are studied
in Section 4.
Finally, our conclusions are presented in the last section.
The discussion is accompanied by explicit examples of
one-dimensional maps.
Non standard definitions used along the text are summarized
in the appendix.

\section{Invariances of fractal dimensions}

Let us consider the map \cite{ott}
\begin{equation}
x_{n+1}=
\left\{
\begin{array}{ll}
2\eta x_n ,& \mbox{if } x_n \leq 1/2, \\
2\eta (x_n -1)+1 ,& \mbox{if } x_n > 1/2, 
\end{array}
\right.
\label{mapa}
\end{equation}
where $\eta$ is a positive parameter ($\eta>1$ in this section).
In spite of its noninversible and discontinuous character, this
map can be seen as a canonical model for hyperbolic scattering
systems. The noninversibility is a consequence of the reduced
dimension of the model. The discontinuity would be avoided
by employing a tend-like map or a logistic-like map, which
is also smooth, with no relevant changes. We prefer this map,
however, because of its simpler symbolic dynamics.

For a uniform distribution of initial conditions in $[0,1]$,
the decay rate with the number $n$ of iterations is $\eta^{-n}$.
A half of the points goes to $x>1$ and then asymptotically to
$+\infty$, and the other half goes to $x<0$ and then to $-\infty$.
In Fig. 1 we show in two different colors the attraction basins
associated to $+\infty$ and $-\infty$ as a function of $\eta$.
In this figure we see the complex structure of the basin boundary
(also present in any magnification), a feature of
systems with sensitive dependence on initial conditions.
The box-counting dimension of the basin boundary
numerically estimated by a method presented in Ref. \cite{ott}
is fitted by $D_c=\ln 2/\ln 2\eta $.
On the other hand,
since the points that go out in one iteration
are those of the interval $(1/2\eta ,1-1/2\eta )$,
we can reinterpret the total exit as being this interval.
In Fig. 2 we show the exit-time function for $\eta =3/2$.
In computing the uncertainty dimension of the
singularities of the exit-time functions by the method of
\cite{lau} we also obtain a curve fitted by
$D_u=\ln 2/\ln 2\eta $.

The coincidence of the above fractal dimensions is due to the
presence of an invariant bounded set (the future
invariant set) with box-counting dimension
$D_c =\ln 2/\ln 2\eta $,
which is chaotic since the Lyapunov exponent $h = \ln 2\eta $
is positive.
The box-counting dimension is defined by \cite{ott}
\begin{equation}
D_c=\lim_{\varepsilon\rightarrow 0}
\frac{\ln N(\varepsilon )}{\ln 1/\varepsilon},
\label{def}
\end{equation}
where $N$ is the number of $\varepsilon$-interval needed to cover
the set.
The invariant set is a Cantor set of Lebesgue
measure zero obtained
as follows. Remove the $1-1/\eta$ open interval
in the middle of
$[0,1]$. From each one of the two remaining
intervals remove the same
fraction in their middle, and so on.

In terms of Fig. 1, the invariant points correspond to the
nonisolated points of the basin boundary for each $\eta$.
A countable set of isolated points, like $x=1/2$, are
in the basin boundary but are not invariant. This set,
refered here as the {\it intrinsic frontier},
is a consequence of
the definition of the basins and, in this
case, do not affect the dimension of the boundary.
(The intrinsic frontier is completely avoided by
coloring in black and white the points that 
leave $[0,1]$ in even and odd number of iterations,
respectively - see Fig. 3.)
Moreover, the basin boundary dimension remains invariant if
we redefine the exits as $(1/2\eta ,\lambda ]$
and $(\lambda , 1-1/2\eta )$
for $1/2\eta <\lambda < 1-1/2\eta $
(Fig. 1 corresponds to the choice $\lambda = 1/2$).
In Fig. 4 we show the basin boundary as a function of
$\lambda $ for $\eta = 3/2$.
In spite of the morphological changes in the
basins, the variation of $\lambda$ just dislocates
the intrinsic frontier so that the basin boundary
dimension is still the dimension of the invariant set.

Let us put the above discussion in a more general
context, where we consider two exits
$E_1$ and $E_2$ so that $E_1\cap E_2={\O}$,
$E_1\cup E_2=(1/2\eta,1-1/2\eta)$,
$\mu (E_1 )\neq 0$ and $\mu (E_2 )\neq 0$
($\mu$ denotes the Lebesgue measure).
The question we address is: How arbitrary can be
the choices (definitions) of the exits
$E_1$ and $E_2$ ?
We denote by $L$ the set of extremes of the intervals
that appear in the construction of the Cantor set,
by $I_0$ the remaining invariant set, by $I$ the
(total) invariant set\footnote{
We call $I_0$ the {\it restricted} invariant set.
The distinction between $I_0$ and $I$ is in order to
show that there is no relevant differences between an
open total exit $(1/2\eta , 1-1/2\eta )$ with
invariant set $I$ and a closed total exit
$[1/2\eta , 1-1/2\eta ]$ with invariant set $I_0$.  
}
$L\cup I_0$,
by $F_1$ the set of points
of the intrinsic frontier in $(1/2\eta , 1-1/2\eta)$
(that go out in the first iteration),
by $F_i$ points of the intrinsic frontier that go
out in the $i^{th}$ iteration, by $F$ the (total)
intrinsic frontier $\cup_{i=1}^{\infty}F_i$,
and by $B$ the (total) basin boundary
$F\cup I$.

Taking the Hausdorff dimension (see \cite{ott} for the definition),
\begin{eqnarray}
D_H(B)&=&\max \{ D_H(F),D_H(I_0),D_H(L)\}\nonumber\\
      &=&\max \{ D_H(F_1),D_H(I_0)\} ,
\label{dh}
\end{eqnarray}
where we used $D_{H}(L)=0$ since $L$ is a
countable set, and $D_H(F)=D_H(F_1)$ because
$F$ is a countable union of similar copies of
$F_1$ \cite{sandau}.
Therefore, the dimension of the basin boundary is equal
to the dimension of the restricted invariant set
if and only if $D_H(F_1)\leq D_H(I_0)$.
When this inequality is not satisfied, the basin
boundary dimension is just an effect of the
definitions of the exits. In the sense of Hausdorff
dimension,
are good exits those ones for which $D_H(F_1)=0$.
It includes all $F_1$ countable, as in the case of
$E_1$ and $E_2$ defined as finite unions of intervals.

Alas, in general we cannot measure $D_H$
but $D_c$, where by definition $D_c(A)\geq D_H(A)$
for any set $A$ \cite{ott}.
Considering the box-counting dimension:
\begin{equation}
D_c(B)=\max\{ D_c(F),D_c(I_0),D_c(L)\}.
\end{equation}
From (\ref{def}) we can show that $D_c(L)=D_c(I_0)$ and
$D_c(F)=\max\{ D_c(F_1),D_c(I_0)\}$.
If $F_1$ is finite, $D_c(F_1)=0$ and
$D_c(B)=D_c(I_0)$.
It is the case for the exits consisting of a
finite union of intervals.
One example is provided by Fig. 4, where $F_1=\{\lambda\}$.
If $F_1$ is infinite, however, $D_c(F_1)$ may be
positive even if $F_1$ is countable (see \cite{ott}, p. 103).
The equality, $D_c(B)=D_c(I_0)$, is satisfied whenever
$D_c(F_1)\leq D_c(I_0)$. In this case we also have
$D_H(B)=D_c(B)=D_H(I_0)=D_c(I_0)$. 
The basin boundary dimension is greater
than the invariant set dimension when $D_c(F_1)>D_c(I_0)$.

In numerical computation, the box-counting dimension is
estimated from the uncertainty dimension ($D_u$) \cite{ott,lau}.
It is conjectured that typically $D_u=D_c$ in dynamical
systems (see \cite{lau} and references therein).
Let us argue about this identity
in our example. First we consider the invariant set
$I$. A point $x_0$ is $\varepsilon$-uncertain if
$(x_0-\varepsilon, x_0+\varepsilon )\cap I\neq {\O}$,
and we denote by $f_I(\varepsilon )$ the fraction of
points that are $\varepsilon$-uncertain.
The dimension of the invariant set is given by
\begin{equation}
D_u(I)=1-\lim_{\varepsilon\rightarrow 0}
\frac{\ln f_I(\varepsilon )}{\ln \varepsilon }\; ,
\label{defdu}
\end{equation}
and it follows from (\ref{def}) that $D_u(I)=D_c(I)$.
Now we consider the boundary set $B$ for $F_1$ finite.
Introducing the exits $E_1$ and $E_2$,
a point $x_0$ is $\varepsilon$-uncertain if $x_0-\varepsilon$
and $x_0+\varepsilon$ go out by different exits.
The fraction of uncertain points is
\begin{eqnarray}
f_B(\varepsilon )
&=&\alpha f_{I} (\varepsilon )+f_{F} (\varepsilon )\nonumber\\
&=&\alpha f_{I} (\varepsilon )+\beta f_{I} (\varepsilon ),
\label{f}
\end{eqnarray}
where $\alpha$ and $\beta$ are constants.
Here, $(1-\alpha)f_I$ is the fraction of points $x_0$
in $f_I$ such that $x_0-\varepsilon$ and $x_0+\varepsilon$
are in the same basin, and it
implies the term $\alpha f_I(\varepsilon)$.
The contribution of the intrinsic frontier is
$f_F(\varepsilon)$, which represents the scale
structure of $I$ for $F_1$. It implies the term
$\beta f_I(\varepsilon )$.
Accordingly,
\begin{eqnarray}
D_u(B)
&=&1-\lim_{\varepsilon\rightarrow 0}
\frac{\ln f_B(\varepsilon )}{\ln {\varepsilon}} \nonumber\\
&=&1-\lim_{\varepsilon\rightarrow 0}
\frac{\ln f_I(\varepsilon )}{\ln {\varepsilon}} \nonumber\\
&=&D_c(I) \;\;(=D_c(I_0)).
\label{du}
\end{eqnarray}
This identity allows the use of
the uncertainty dimension, which has nice
numerical properties, to calculate the
box-counting dimension.

Back to Fig. 2, the set of singularities ($S$) of the
exit-time function (set of point where the exit-time
is infinite) corresponds to the invariant set $I$.
For the
exit-time $T$, a point $x_0$ is $\varepsilon$-uncertain
if $T(x_0-\varepsilon)\neq T(x_0+\varepsilon)$. Therefore,
in the computation of the uncertainty dimension we are in
fact estimating the box-counting dimension
of the discontinuities of the exit-time
function. (In this case there is no intrinsic frontier,
since the exit-time is defined with respect to
the total exit.) The set of discontinuities can be
identified with $L$, forming sequences that converge to the
singularities. It implies
\begin{equation}
D_u(S)=D_c(L)=D_c (I).
\label{du2}
\end{equation}
Once more the uncertainty dimension is
the same as the box-counting dimension.

Although discussed for a specific case, the above results
are far more general since our system models
a vast class of exit systems.
For any system with a finite number of basins,
the dimension of the union of all basin boundaries
is equal to the dimension of
at least one of the boundaries.
So, it is enough to consider systems with only one
boundary (two exits, $E_1$ and $E_2$.)
When the effects of the intrinsic frontier are
avoided, the basin boundary has the same
dimension as the future invariant set.
Since the latter set does not depend on $E_1$ or $E_2$
but just on $E_T=E_1\cup E_2$,
it leads to an invariance of the dimension
with the definition of the basins.
Roughly speaking, 
in spite of possible morphological changes,
the basin boundary dimension
will be the same for any
nonpathological couple of exits $E'_1$ and $E'_2$
with positive probability of being reached,
even since satisfying $E_T=E'_1\cup E'_2$.
Then, there is usually great freedom
in defining the exits for a fixed total exit.
This invariance of fractal dimension allows us to
choose among several alternatives the best one to measure
the invariant set dimension and hence the sensitivity
of the system.
Also, it gives us confidence that the result
reflect a true dynamical property of the system
and not an immaterial effect.

In this section we studied measurements of
fractal dimensions
when the total exit is fixed and, therefore,
the invariant set is fixed. Next, in Sections 3 and 4,
we discuss the complementary situation, that is,
variations of the invariant set (and of its dimension)
under changes of the total exit.
 
\section{Dependences with the total exit}

A fundamental feature of map (\ref{mapa})
is the monotonous increase of the (future) invariant set
dimension with the reduction of the
parameter $\eta$.
For $\eta >1$, the invariant set is a zero measure
set satisfying $D(I)\rightarrow 1$ for $\eta\rightarrow 1$.
This property is very general and closely related to
a conjecture of Moura and Letelier \cite{moura}
in the context of Hamiltonian systems:
The dimension of the invariant set of a chaotic
scattering system goes to the effective phase space
dimension when the energy tends to the escape energy.
In both cases the reduction of a parameter, $\eta$ or
the energy, blinds the exits and increases the
number of oscillations of typical orbits before
going out. Accordingly, it ``fattens'' the invariant set,
increasing its dimension.

Let us consider an extension of the
above situation to one where the total exit is enlarged.
Instead of $(1/2\eta , 1-1/2\eta )$ (the natural exit
determined by points that go to $\pm\infty$) we define
the new total exit $(1/2\eta-\alpha , 1-1/2\eta+\alpha )$
for $0<\alpha<1/2\eta $ and $\eta >1$.
The invariant set $I(\eta,\alpha)$ will be
in $[0,1/2\eta-\alpha]\cup[1-1/2\eta+\alpha,1]$
and its dimension can be easily computed from the
exit-time function. In Fig. 5 we show 
the dimension of the invariant set as a function
of $\alpha$ for $\eta =1.1$.
The fractal dimension is a decreasing function of
$\alpha$, but {\it not} strictly decreasing.
The dimension presents a pattern of plateaux due to the
existence of gaps in the invariant set.
A gap $(1/2\eta -\alpha_2,1/2\eta-\alpha_1)
\cup(1-1/2\eta+\alpha_1,1-1/2\eta+\alpha_2)$
in the invariant set $I(\eta,\alpha_1)$ implies a
plateau in the interval $(\alpha_1,\alpha_2)$: In
increasing $\alpha$ the dimension changes only when
the enlarged exit advances on points of the invariant
set $I(\eta,\alpha)$.
The result is a graph with a staircase structure where
the derivative is null in the complement of a Cantor set.
Moreover, the limit of small $\alpha$ is well
defined and recover the original dimension, that is,
$D(I(\eta,\alpha)) \rightarrow D(I(\eta,0))=\ln 2/\ln 2\eta$
for $\alpha\rightarrow 0$.

It is instructive to consider the invariant set explicitly
in terms of its symbolic dynamics. The symbolic dynamics
can be associated to the invariant set of map (\ref{mapa})
as follows. For the exit $(1/2\eta ,1-1/2\eta )$,
an invariant point $x_0\in [0,1]$ is represented by
$0.\epsilon_0\epsilon_1. . . \epsilon_n . . . \; ,$
where $\epsilon_n=0$ if $M^n(x_0)\leq 1/2\eta$
and $\epsilon_n=1$ if $M^n(x_0)\geq 1-1/2\eta$.
The grammar of this dynamics is the Bernoulli shift grammar
and $M^n(x_0)$ is represented by
$0.\epsilon_{n}\epsilon_{n+1} . . . \; .$
To simplify, suppose the exit is enlarged only
on the left becoming $(1/2\eta -\beta ,1-1/2\eta)$,
where $\beta$ is so that points in
$(1/2\eta-\beta ,1/2\eta)$
are represented by
$0.011...1\epsilon_{m}\epsilon_{m+1}...$
for some integer $m$.
The points of the invariant set with respect to the
exit $(1/2\eta ,1-1/2\eta )$ that go out in fewer than
$k\times m$ iterations are those points represented
by sequences of $0$ and $1$ with the word
$0\underbrace{11...1}_{m-1}$ somewhere in the first
$k\times m$ letters.
Subtracting these points, the fraction of invariant
points with respect to $(1/2\eta ,1-1/2\eta )$
that remain invariant with respect to the exit
$(1/2\eta-\beta ,1-1/2\eta)$
for $k\times n$ iterations is
\begin{equation}
N(k,m)=2^{km}\left\{
1-\sum_{s=1}^{k}
\left[
\frac{(-1)^{s+1}}{2^{sm}}\sum_{i_s}^{(k-s)n+1}...
\sum_{i_2}^{i_3}\sum_{i_1}^{i_2}1
\right]
\right\}.
\label{km}
\end{equation}
It gives the scale of the hierarchical construction of the
Cantor set structure,
from which follows that the box-counting dimension
of the new invariant set is
\begin{equation}
D_c(I(\eta,\beta))=\lim_{k\rightarrow\infty}
\frac{\ln N(k,m)}{\ln (2\eta )^{km}} \; .
\label{geral}
\end{equation}
For $m=3$, the computation of the sequences in a
different order leads to
\begin{equation}
N(k,3)=2^{3k-1}
\left\{
\sum_{j=2}^{3k}a_j+\sum_{j=3}^{3k}b_j
\right\},
\label{k3}
\end{equation}
where $a_2=b_3=1$,
\begin{equation}
a_j=-\frac{f_{j-2}}{2f_j}
\sum_{i=2}^{j-1}a_i\;\mbox{for}\; j\geq 3,\;
b_j=-\frac{f_{j-1}-1}{2f_{j+1}-2}
\sum_{i=3}^{j-1}b_i\;\mbox{for}\; j\geq 4,
\label{ab}
\end{equation}
and $\{f_n\}$ are the Fibonacci numbers.
The case $m=2$ is trivial since $N(k,2)=2k+1$
and the dimension of the invariant set is zero.
The dimension computed from (\ref{km}-\ref{ab})
is identical to that one computed
by the uncertainty method. In Fig. 6 we show this
fractal dimension, normalized by the factor
$\ln 2/\ln 2\eta$, as a function of $m$.
The exit size is reduced as $m$ is augmented,
increasing the dimension.

The enlargement of exits is useful and sometimes
necessary in numerical computation of fractal
dimensions.
An example is given by the orbits of light
traveling around a black hole, as discussed in
Section 1. The difficulty in determining the
final state (at the black hole or at infinity)
is removed by enlarging the exit at infinity.
The new exit may be defined as $r>r_0$, instead
of $r=\infty$, where $r$ is an adequate distance
from the attractor center. The dimension so
obtained is an under bound of the exact
invariant set dimension (associated to the
original exits) since part of the invariant set
is lost throughout the new exit.

The difficulty illustrated here is present in a vast
class of systems. It has frequently appeared
in the literature where, in general, some kind of
enlargement of the exits is implicitly taken into account. 
The above discussion intends to provide subsidies
to clarify the effects of this procedure.

\section{Transitions from exit to nonexit systems}

Exits can be defined even when the system has no natural
exits. As an example consider map (\ref{mapa})
once more, but now for $\eta =1$
(it is analogous for $1/2<\eta <1$).
In such case the interval [0,1] is mapped to itself,
and the symmetrical exit defined in Section 3
reduces to $(1/2-\alpha, 1/2+\alpha)$. In Fig. 7 we show
a graph of the dimension of the (zero measure)
invariant set as a function of $\alpha$.
The graph presents a pattern of plateaux
similar to that of Fig. 5.
In fact, it is a consequence of a simple
relation existing between fractal dimensions
for $\eta>1$ and for $\eta =1$.
This relation follows from a one-to-one correspondence,
via symbolic dynamics\footnote{
For $\eta=1$, all the interval $[0,1]$ is associated to the
Bernoulli shift symbolic dynamics. The picture is
analogous to that one for $\eta>1$, except that
the original invariant set (for $\alpha =0$)
has measure zero for $\eta>1$ and measure one for
$\eta=1$.
},
between invariant sets for
different values of $\eta$.
In terms of symbolic dynamics, the exit
$(1/2\eta-\alpha, 1-1/2\eta+\alpha)$ can be denoted by
$E=(0.0\epsilon_1\epsilon_2...\;
,\; 0.1\epsilon'_1\epsilon'_2...)$,
where $\epsilon_i,\epsilon'_i\in\{0,1\}$ and
$\epsilon_i+\epsilon'_i=1$.
This notation ignores the plateaux
due to the original exit since
$0.0\epsilon_1\epsilon_2...=\psi_{\eta}(\alpha)$
where the function $\psi_{\eta}$ is constant on
gaps of the original invariant set. Accordingly, from
(\ref{def}) follows that the box-counting dimensions
for $\eta=1$ and $\eta>1$ differ just by the overall
scale factor $\ln 2/\ln 2\eta$,
\begin{equation}
D_c(I(\eta,E)) =
\frac{\ln 2}{\ln 2\eta }D_c(I(1,E)).
\label{e1}
\end{equation}
(Relation (\ref{e1}) is valid for
any exit, not only for the one considered here,
e.g. Fig. 6 where $d=D_c$ for $\eta=1$.)

An important feature exhibited in Fig. 7 is that
the dimension is fractal and goes to $1$ as
$\alpha$ goes to $0$. This behavior
is closely related to the fact that the original map,
with no exit, is chaotic: the map has a dense
set of unstable periodic orbits and the Lyapunov
exponent is positive ($h=\ln 2$). In order to better
understand this relation, let us consider the
logistic map $x_{n+1}=rx_n(1-x_n)$ in the unit
interval $[0,1]$ for $r=3.8$ (chaotic)
and for $r=2.8$ (nonchaotic).
In Fig. 8 we plot the exit-time function
for $r=3.8$ with respect to the exits
$(0.61,0.69)$ (Fig. 8a) and
$(0.63,0.67)$ (Fig. 8b).
These graphics present successive steps
in the construction of the Cantor structure
of the invariant set.
They must be compared with Fig. 9,
where we show exit-times for $r=2.8$.
The exit-time for $r=3.8$ presents a
complex structure that becomes even more complex
when the exit is reduced. It reflects
the increasing of the invariant
set dimension, $D_u=0.62$ in Fig. 8a, $D_u=0.88$ in Fig. 8b,
and $D_u\rightarrow 1$ in the limit of small exit.
A different behavior takes place when $r=2.8$.
In fact, Figs. 9a and 9b show a regular exit-time
where no complexity is added when the exit is reduced.
In terms of invariant set dimension 
it means that its value is zero even in the limit
of small exit.
The transition of the dimension of the invariant set
from small exit to no exit is, in this case,
discontinuous since it jumps from $0$ to $1$.

The behavior described in preceding
paragraphs is
very general and may be summarized as follows:
Chaotic nonexit systems present invariant
sets with {\it fractal} dimension when small
exits are defined\footnote{
The exits have to be sufficiently small
in order to avoid the complete outcome of
the invariant set.
},
and the invariant set dimension
tends to the effective phase space dimension
when the exits are arbitrarily reduced.
On the other hand, nonchaotic nonexit
systems do not present any
fractal structure in the invariant set when exits
are introduced, and the invariant set dimension
jumps (discontinuously) to the effective phase
space dimension when the exits are removed\footnote{
There are pathological examples of nonchaotic systems
which exhibit fractal properties
when exits are created (see \cite{troll}).
It happens when the invariant set of the
exit system is fractal but nonchaotic.
This behavior is, however, atypical.
}.
We conjecture that this behavior is typical
for dynamical systems in general.

Basically, this conjecture states that chaos
in closed systems and fractals in open systems
are different manifestations of the same phenomenon.
Chaos in closed systems is essentially determined
by the presence of a somewhere dense set of
unstable periodic orbits. Fractal
structures in exit systems are determined by a fractal
invariant set that consists of unstable periodic orbits
and an uncountable number of nonperiodic orbits
surrounding them. Therefore, the introduction and
removal of exits leads from one situation to the other
(In most cases, because of the density property, the
box-counting dimension of the periodic orbits equals the
future invariant set dimension in exit systems and the
effective phase space dimension in closed systems.)
Physically, an exit system with fractal invariant set
evolves chaotically for a period of time
before been scattered. When the exits are removed the
system evolves chaotically forever.
Conversely, chaotic nonexit systems present
sensitive dependence on initial conditions.
The introduction of small exits
detect this sensitivity through singularities on
exit-times, scattering functions, etc.

It is also opportune to consider
a couple of related works,
\cite{bleher} and \cite{mot}.
In the former, the authors studied the Sinai
billiard with two holes in the external wall.
They show that the boundary between the
initial-condition basins of attraction defined
by the two exits is fractal.
In the context of our discussion,
this result is a consequence of the well
known chaotic dynamics of the (closed)
Sinai billiard. In addition, we verified
numerically that the basin boundary dimension
in fact tends to the effective phase space dimension
when the two exits are reduced.
In the latter work, the question
over whether or not the Bianchi IX cosmological model
is chaotic was finally solved.
The trouble with this question was that standard
indicators of chaos, like Lyapunov exponents,
are not invariant under space-{\it time} diffeomorphisms
and cannot be used since the model is relativistic.
Following a suggestion first made by \cite{cor},
the authors show that the system is chaotic
by defining three exits and measuring the basin boundary
dimension. (The approximation of the dimension to the
effective phase space dimension when the exits are
reduced was also verified.) In that work, the relation
between exit and nonexit systems together with the
invariant character of fractal methods played a
fundamental role in the solution of the problem.

The potential use of our conjecture as an
invariant method to study chaos in closed systems
of several dimensions is considered in \cite{pre}.

\section{Conclusions}

We have studied, via examples,
a number of questions concerning fractals
in exit systems.
In particular the three questions stated
in Section 1 were discussed in Sections 2, 3 and 4,
respectively.
We observed that basin boundary
dimensions are independent of the definitions of
the exits since the total exit is fixed and the
intrinsic frontier effects are avoided.
The dimensions increase, however, with the reduction
of the total exit. Our main result refers to
the limit of small exits, where a distinctive behavior
was found concerning whether the corresponding closed
system is chaotic or not. In the former the dimension
changes continuously, while in the latter it jumps
avoiding noninteger values.
Special attention was also given to understand
the relations
between the dimensions of different fractal sets, and
the relations between different concepts of dimension. 

Finally, we stress that most of the results presented
here are also valid for nonhyperbolic systems \cite{lau}.
It is the case of the invariances of fractal dimensions
discussed in Section 2, since the arbitrariness
in choosing the exits is analogous in hyperbolic and
nonhyperbolic systems.
Concerning Section 3, however,
the dimension changes with the total
exit only when the nonhyperbolic character is lost in enlarging
the exit. In such case, the behavior of hyperbolic systems
discussed in Section 3 may take place. Nonhyperbolic exit systems
are supposed to have invariant set with maximal dimension
(equal to the dimension of the effective phase space) \cite{lau}.
In this sense, and in the context of Section 4, the relation
between closed and open nonhyperbolic systems is even stronger
than in the hyperbolic analogous.

\vskip 0.5truecm

\leftline{ACKNOWLEDGMENTS}

\vskip 0.2truecm

\noindent The authors thank Fapesp and CNPq for financial support.

\vskip 0.5truecm

\leftline{APPENDIX}

\vskip 0.2truecm

\noindent The following definition refers to phase spaces of
autonomous systems for both maps (discrete time) and vector
fields (continuous time):
(1) An {\it exit} $E$ is any set of the phase space (possibly at
infinity) which may be reached in a finite time or asymptotically.
(2) The {\it total exit} $E_T$ is the union of all exits of the
system.
(3) The {\it size} of an exit is given by its phase space volume,
and an exit $E$ is {\it smaller} than another exit $E'$
if $E$ is a proper subset of $E'$. The size of the total exit is
defined analogously.
(4) Given a set of exits $E_1, E_2,...,E_n$,
the {\it basin of attraction}
$A_i$ of the exit $E_i$ is the set of all initial
conditions whose time evolution reaches
$E_i$ before reaching any other
exit\footnote{
$A_i$ would be defined as the closure of this
set, with no relevant changes.
}.
(5) The {\it basin boundary} $B_{{i_1},...,{i_k}}$ is the set
of points arbitrarily close to points of
$A_{i_1}$,..., $A_{i_{k-1}}$, and $A_{i_k}$.
(6) A {\it scattering function} is a function that describes the
behavior of the final phase space variables
(when the exits are reached) as
a function of the initial ones.
(7) The {\it exit-time function} (``time delay'' in scattering
processes) is a function which associates to the initial conditions
the minimal time needed to reach the total exit.
(8) The {\it nonattracting invariant set}
is a set composed by nonattracting
orbits that do not reach any exit
(not even asymptotically)
for both directions of the time.
Its {\it stable} and {\it unstable manifolds},
when exist, are the future and past invariant sets,
respectively.

\newpage

\begin{figure}
\caption{Portrait of the basins of map (1)
as a function $\eta$.
The initial conditions were chosen on a grid of $400\times 400$.
Regions in black and white correspond to orbits that escape to
$+\infty$ and $-\infty$, respectively.
}
\label{fig1}
\end{figure}

\begin{figure}
\caption{Exit-time of map (1) for $\eta =3/2$
and total exit $(1/2\eta ,1-1/2\eta )$.
}
\label{fig2}
\end{figure}

\begin{figure}
\caption{Portrait of the basins of map (1)
as a function $\eta$.
The initial conditions were chosen on a grid of $400\times 400$.
Regions in black and white correspond to orbits that escape 
from $[0,1]$ in an even and an odd number of iterations,
respectively.
}
\label{fig3}
\end{figure}

\begin{figure}
\caption{Portrait of the basins for map (1)
as a function of $\lambda $ for $\eta =3/2$.
The initial conditions were chosen on a grid of $800\times 800$.
Regions in black and white correspond to orbits that escape to
$(1/2\eta ,\lambda ]$
and $(\lambda , 1-1/2\eta )$,
respectively.
}
\label{fig4}
\end{figure}

\begin{figure}
\caption{Uncertainty dimension of the invariant set
of map (1) for $\eta=1.1$ and total exit
$(1/2\eta-\alpha , 1-1/2\eta+\alpha )$
as a function of $\alpha $.
}
\label{fig5}
\end{figure}

\begin{figure}
\caption{The normalized box-counting dimension
$d=D_c\ln 2\eta/\ln 2$
of the invariant set of map (1) for the total exit
$(1/2\eta-\beta, 1-1/2\eta)$
as a function of $m$.
}
\label{fig6}
\end{figure}

\begin{figure}
\caption{Uncertainty dimension of the invariant set
of map (1)
for $\eta=1$ and total exit $(1/2-\alpha , 1/2+\alpha )$
as a function of $\alpha $.
}
\label{fig7}
\end{figure}

\begin{figure}
\caption{Exit-time of the logistic map for $r=3.8$
with total exit:
(a) $(0.61,0.69)$; (b) $(0.63,0.67)$.  
}
\label{fig8}
\end{figure}

\begin{figure}
\caption{Exit-time of the logistic map for $r=2.8$
with total exit:
(a) $(0.61,0.69)$; (b) $(0.63,0.67)$. 
}
\label{fig9}
\end{figure}

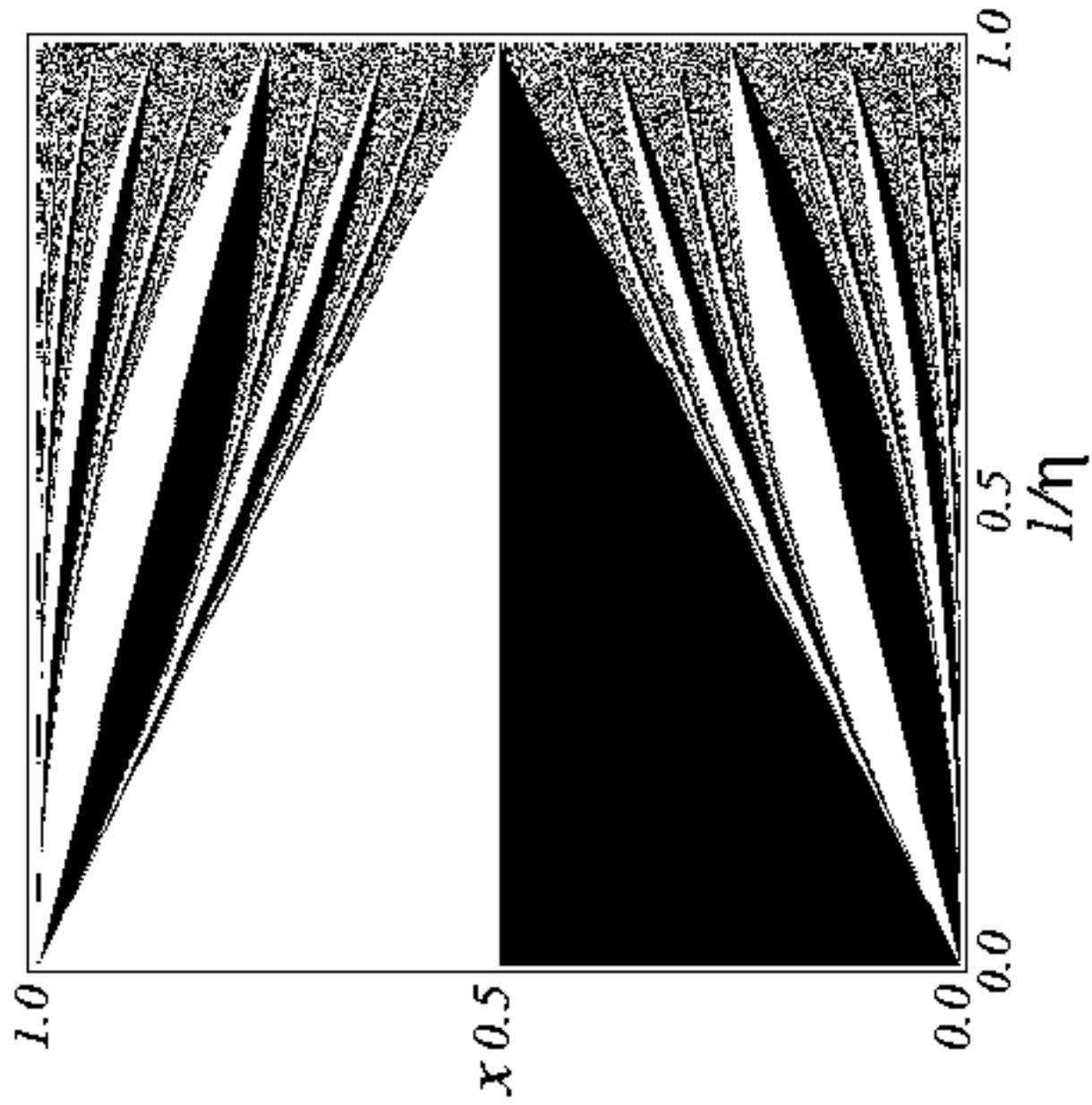
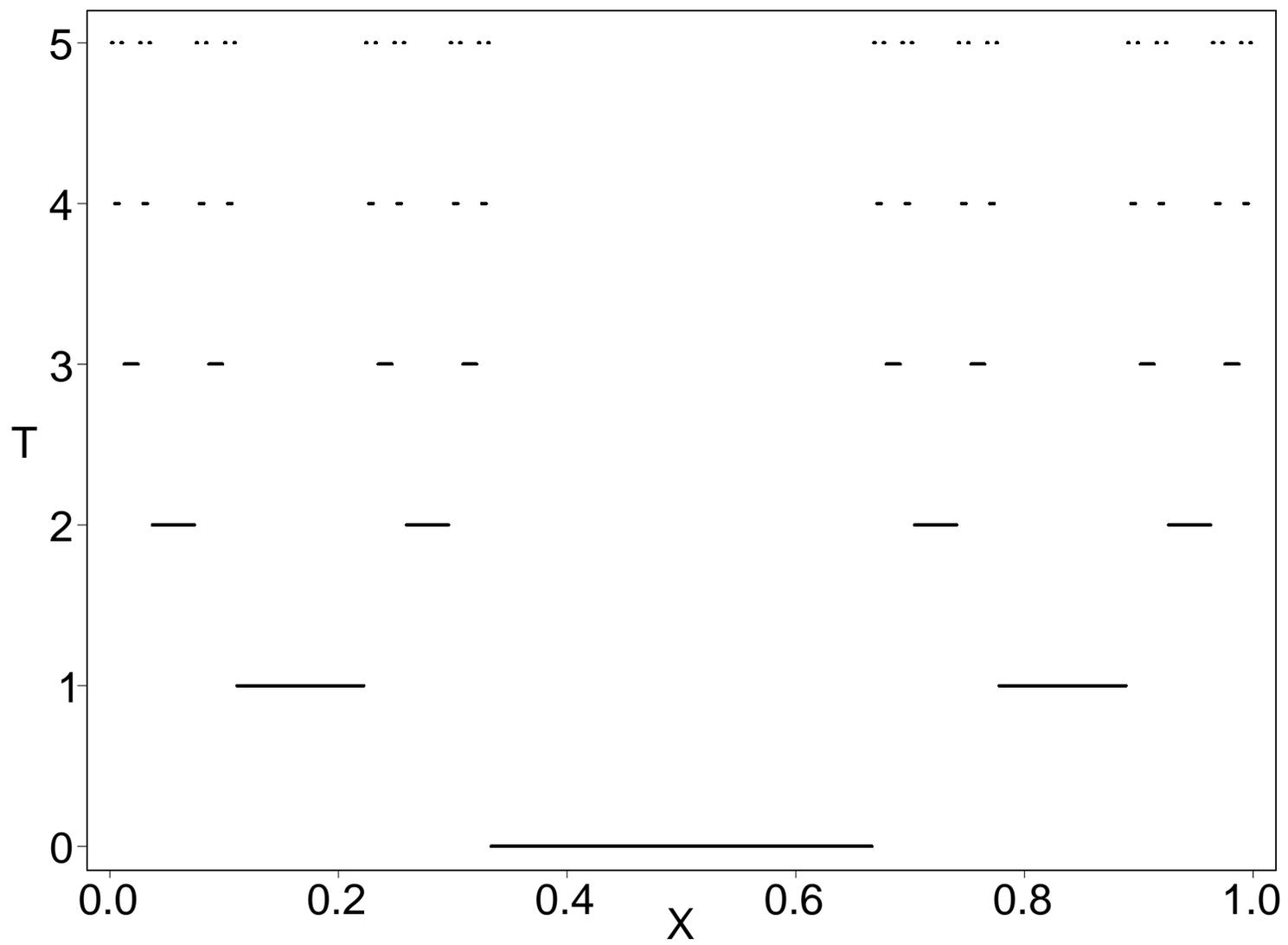
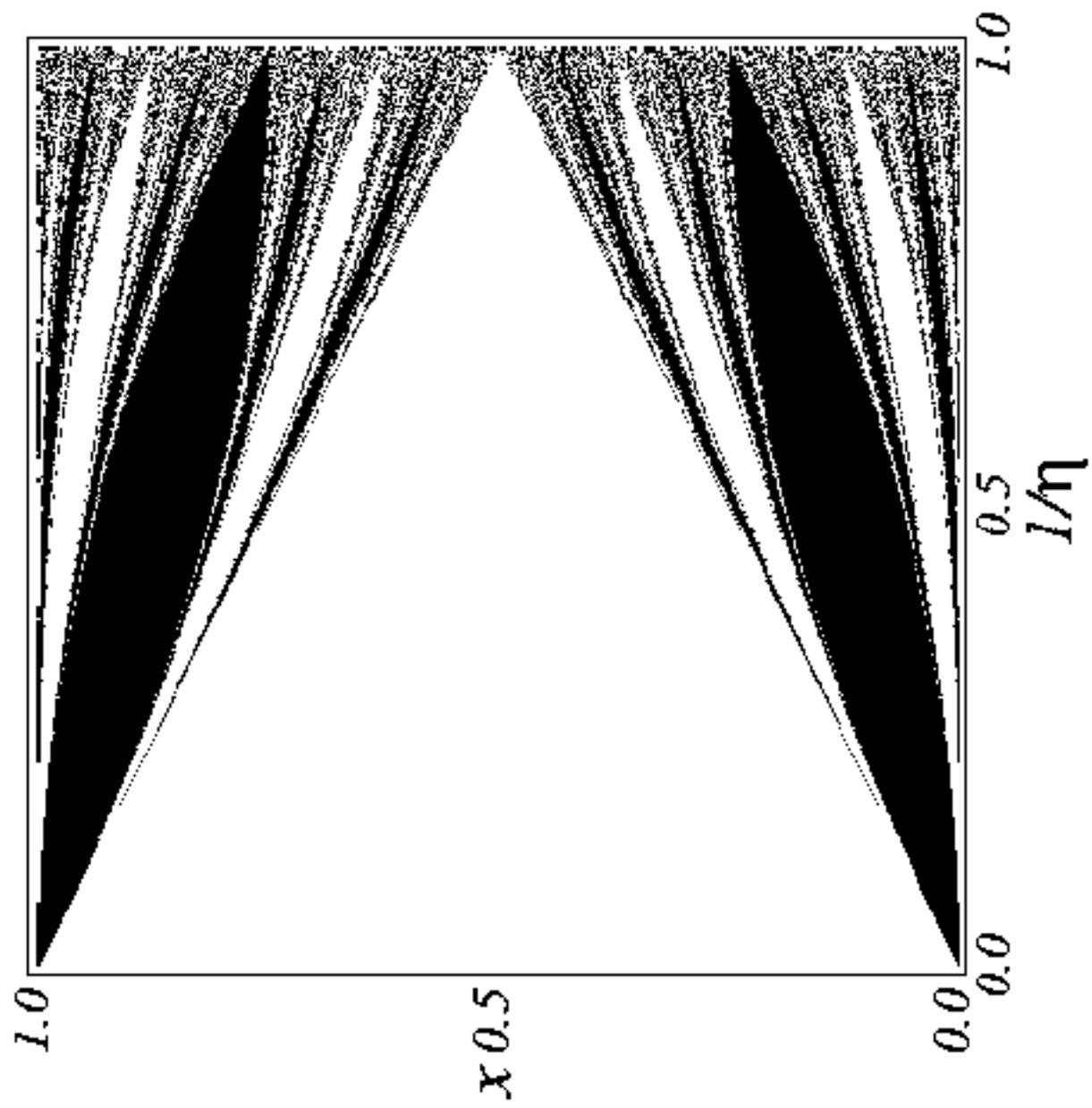
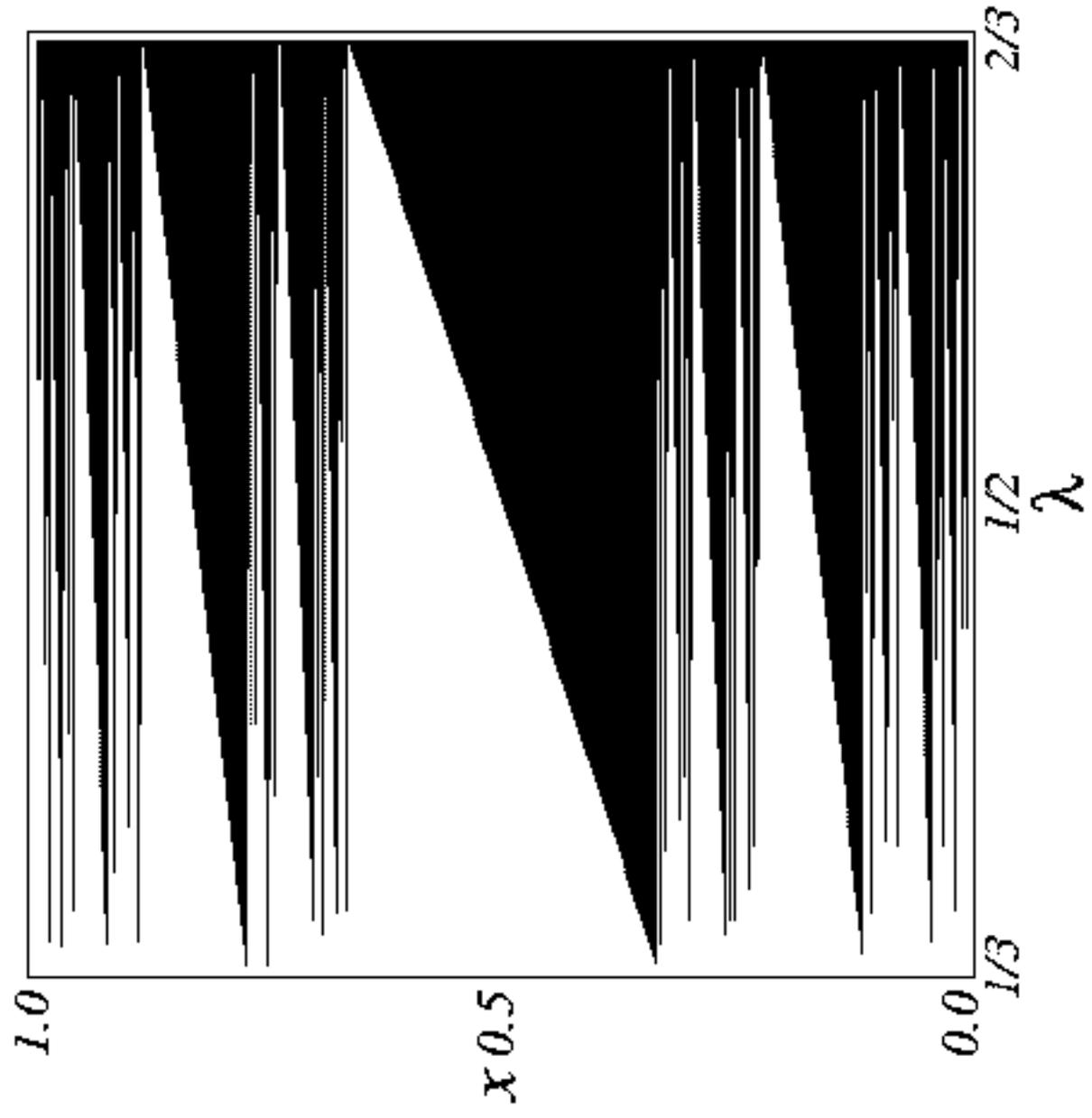
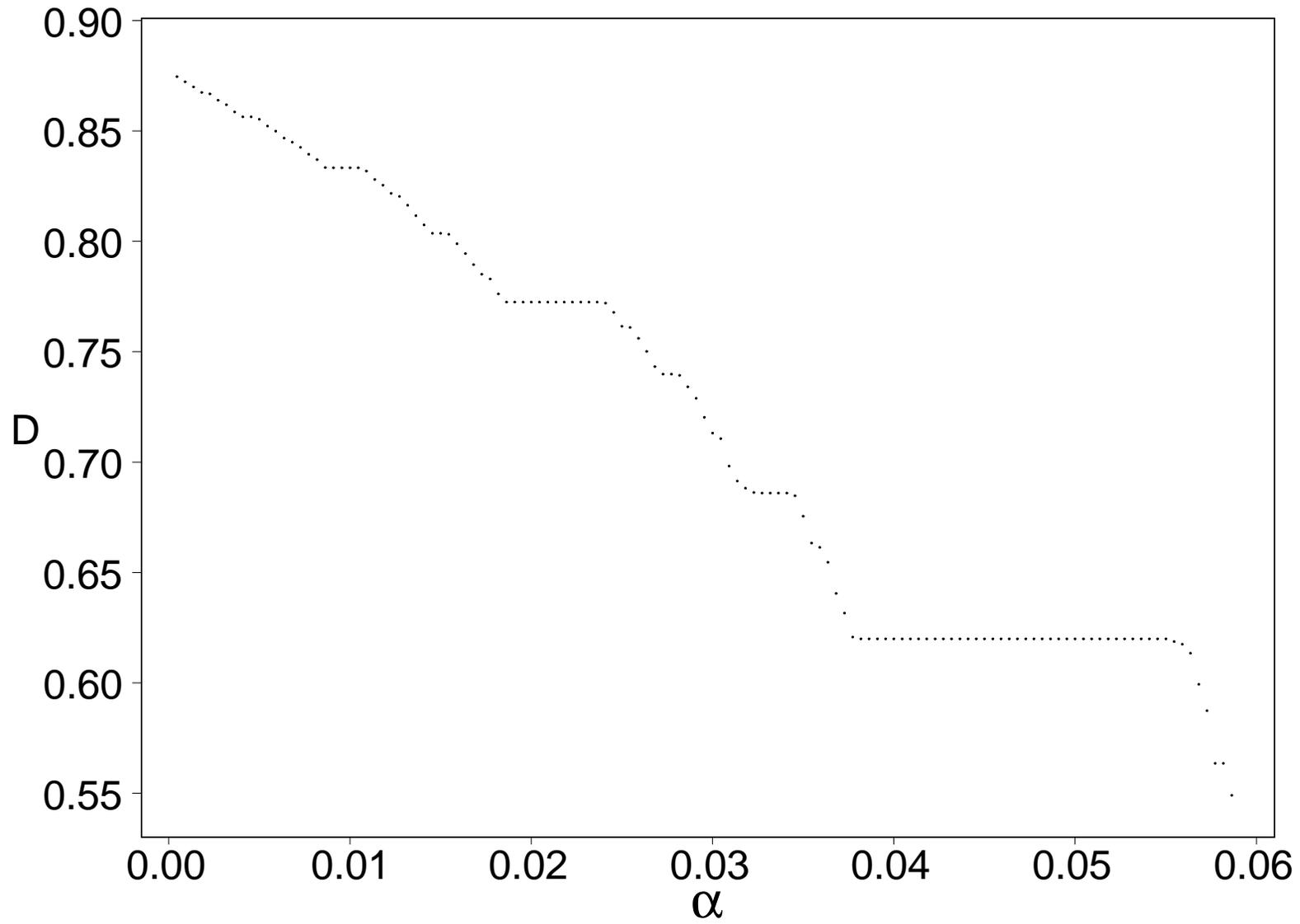
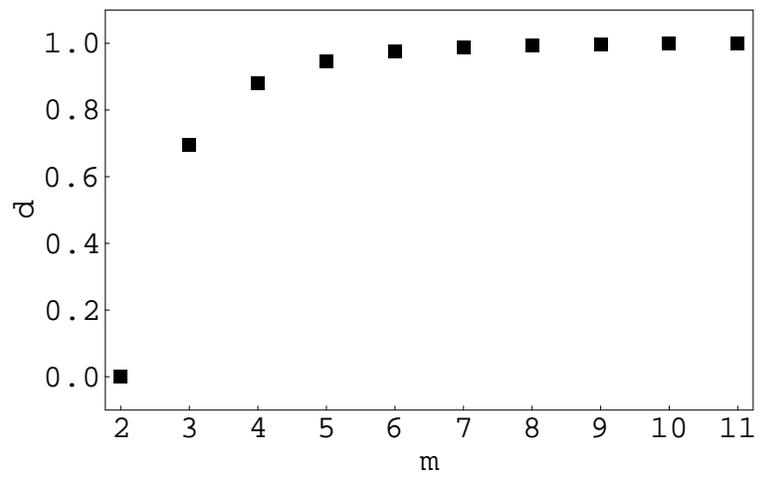
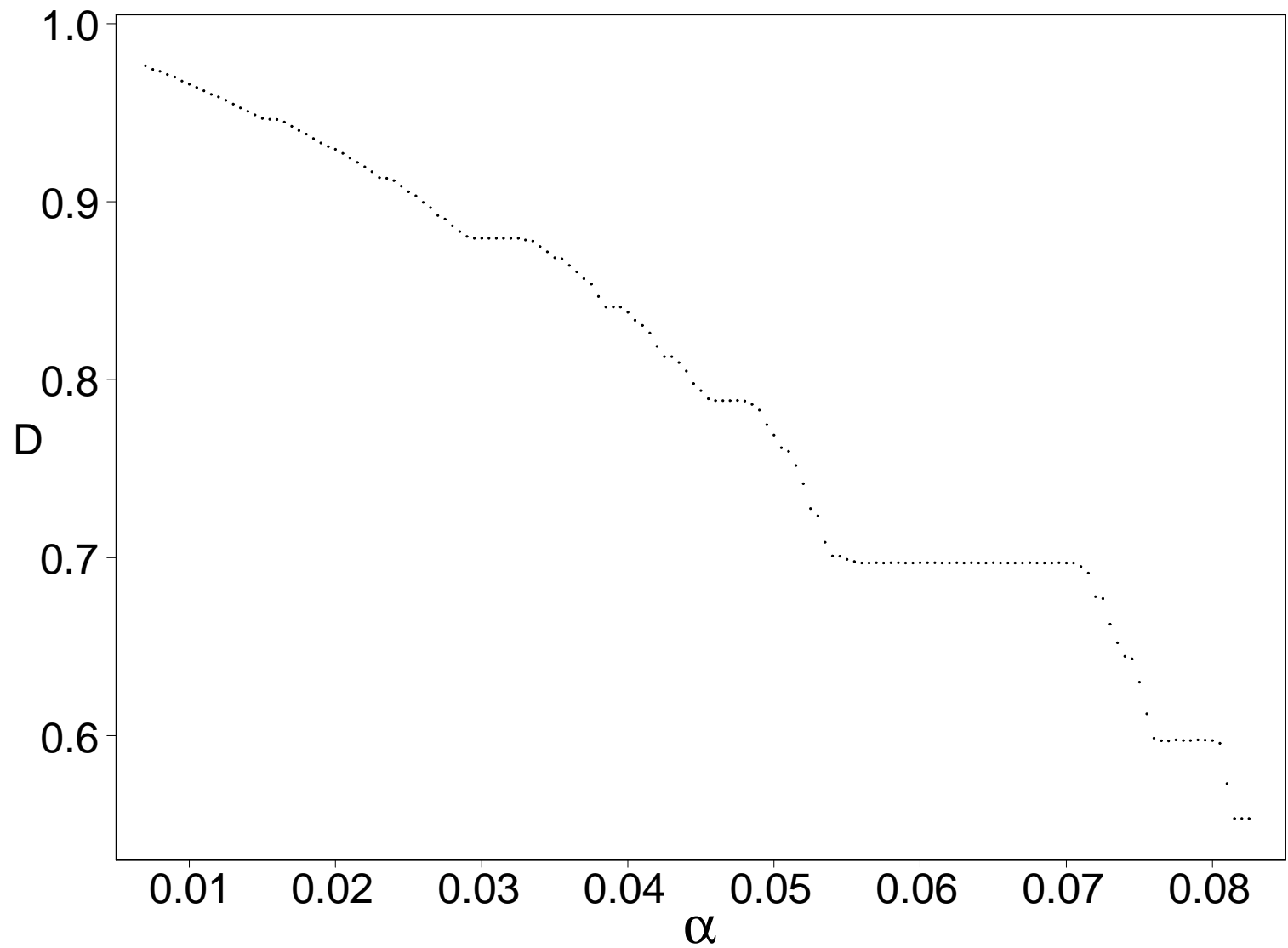
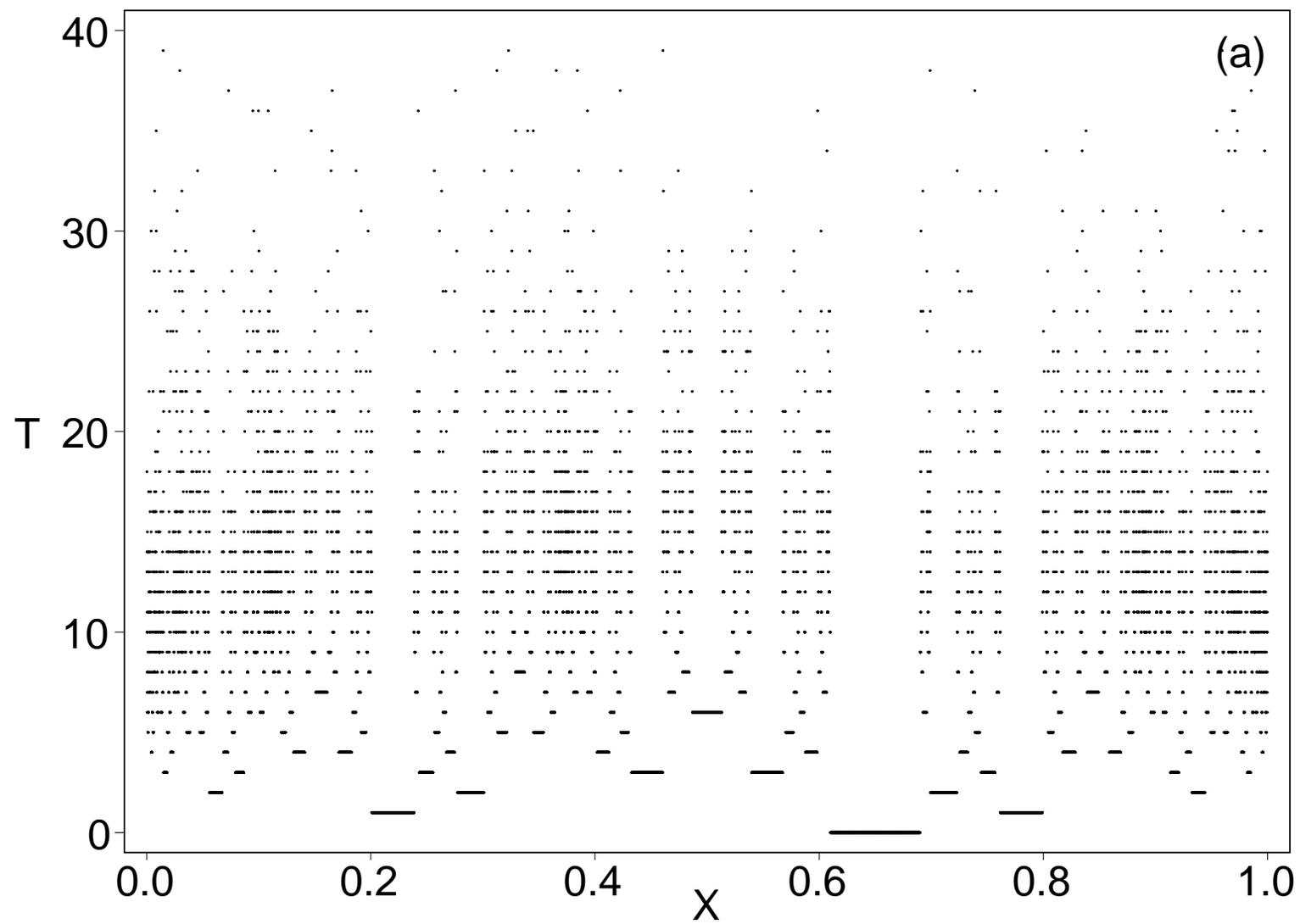
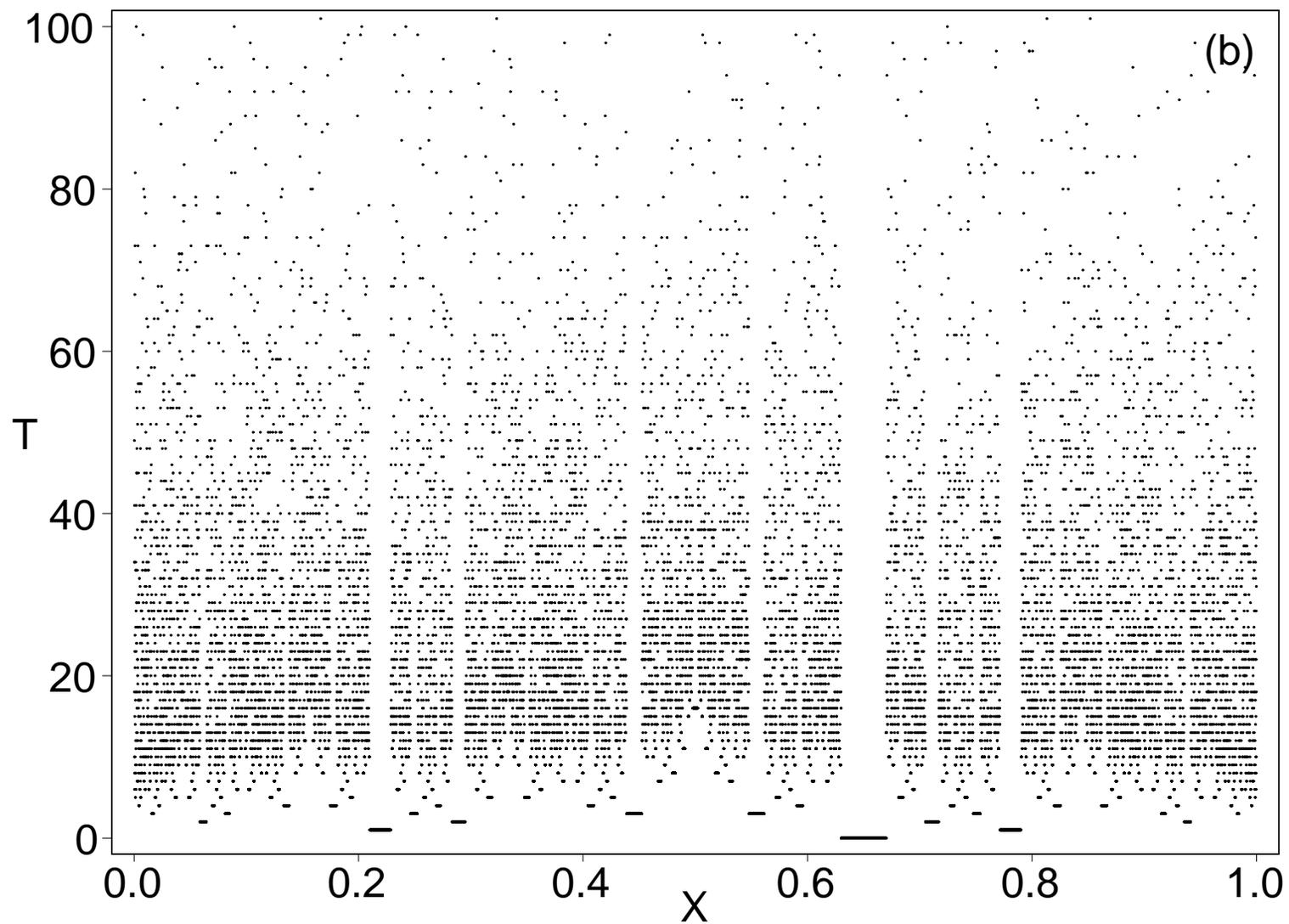
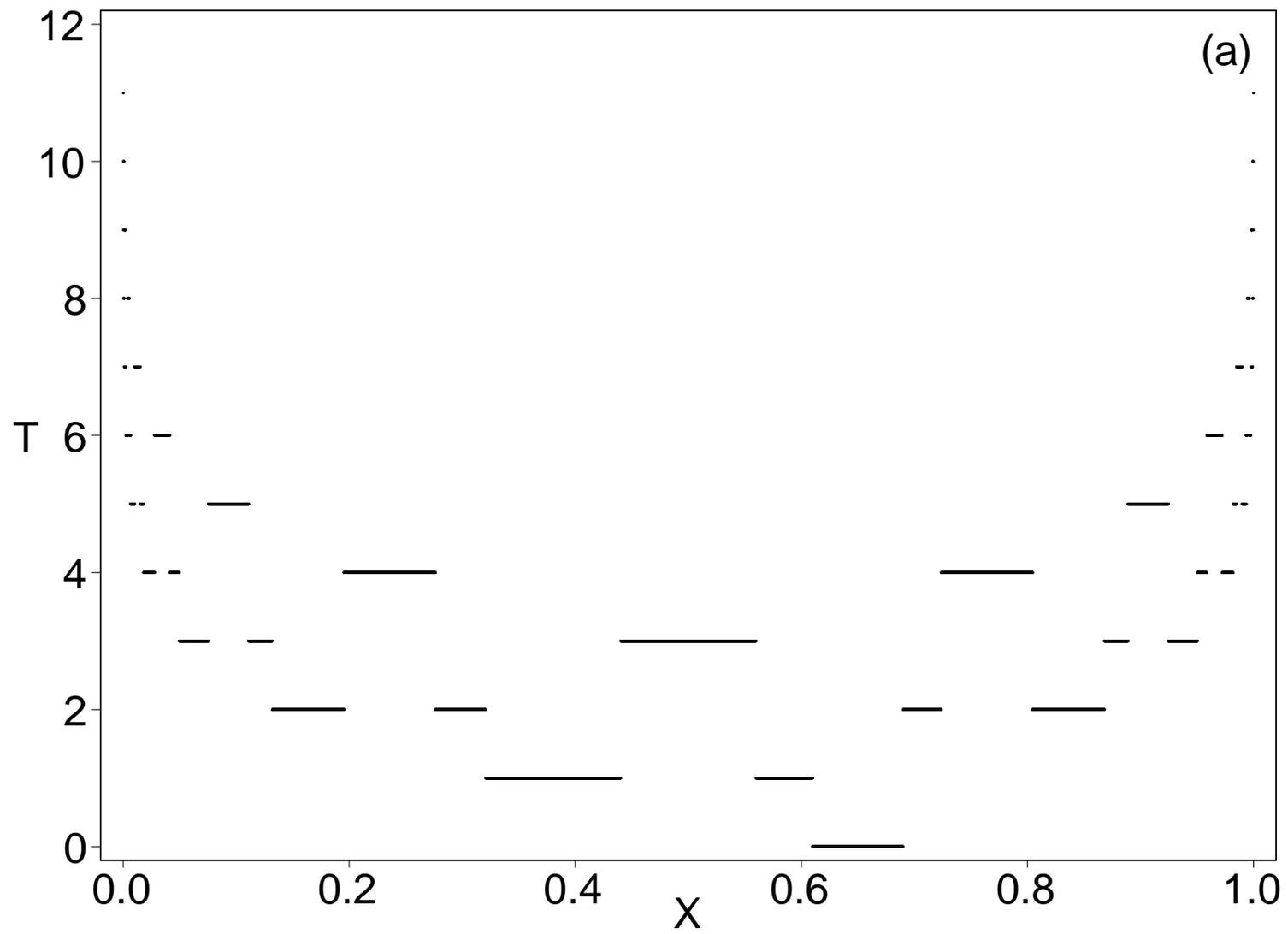
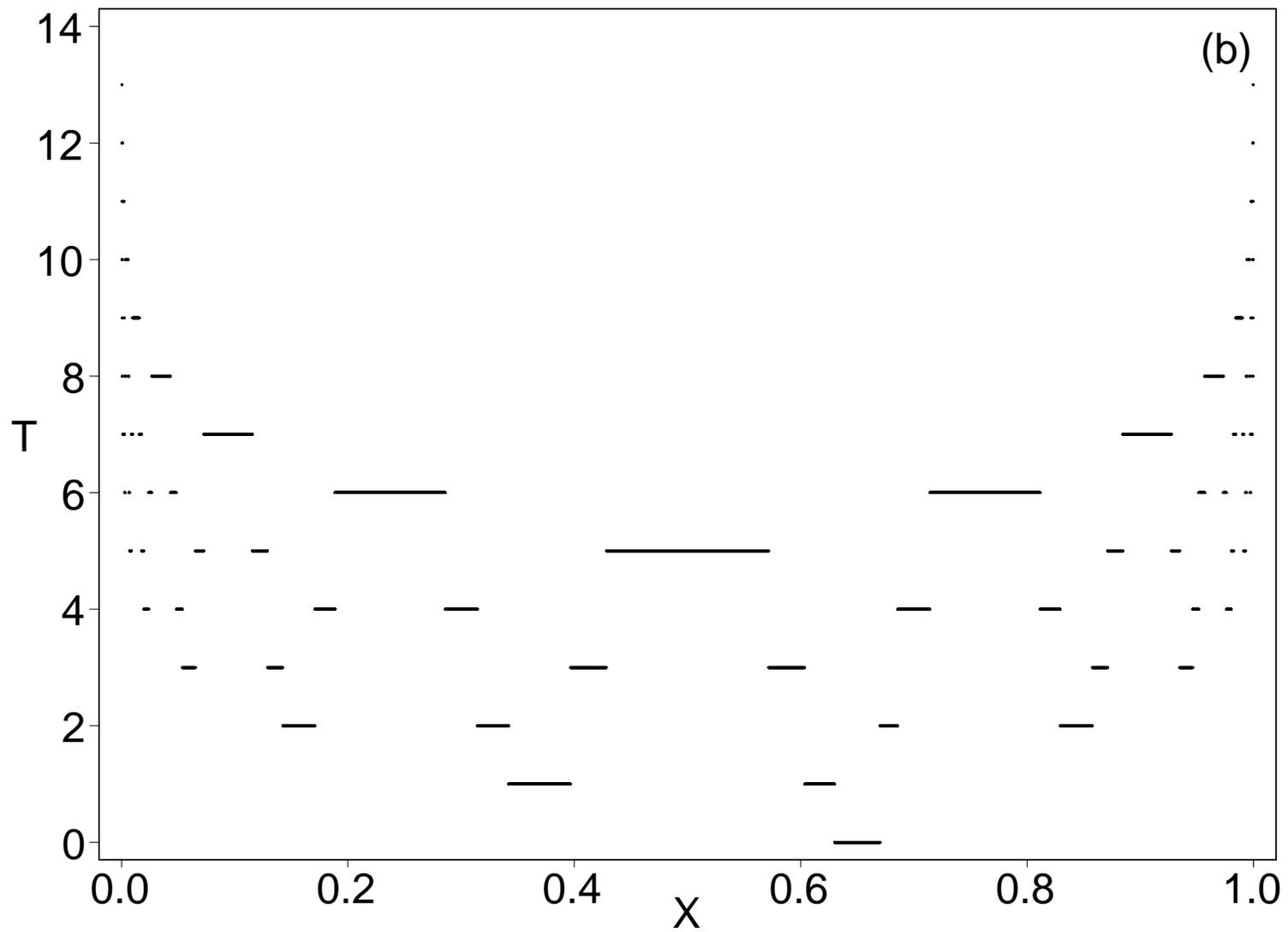
\end{document}